\documentclass[conference]{IEEEtran}
\IEEEoverridecommandlockouts
\usepackage{cite}
\usepackage{amsmath,amssymb,amsfonts}
\usepackage{algorithmic}
\usepackage{graphicx}
\usepackage{textcomp}
\usepackage{xcolor}
\definecolor{lightgray204}{RGB}{204,204,204}
\usepackage{comment}
\usepackage{soul}
\sethlcolor{red}
\usepackage{url}
\usepackage[acronym,shortcuts,nonumberlist,nopostdot,nomain,nogroupskip]{glossaries}
\newacronym{CSI-RS}{CSI-RS}{channel state information-reference signal}
\newacronym{SRS}{SRS}{sounding reference signal}
\newacronym{P2P}{P2P}{point to point}
\newacronym{UPA}{UPA}{uniform planar array}
\newacronym{ULA}{ULA}{uniform linear array}
\newacronym{PMI}{PMI}{precoding matrix indicator}
\newacronym{RI}{RI}{rank indicator}
\newacronym{CQI}{CQI}{channel quality indicator}
\newacronym{DL}{DL}{downlink}
\newacronym{CDF}{CDF}{cumulative distribution function}
\newacronym{DFT}{DFT}{discrete Fourier transform}
\newacronym{2D-DFT}{2D-DFT}{two dimensional discrete Fourier transform}
\newacronym{NR}{NR}{new radio}
\newacronym{LTE}{LTE}{long term evolution}
\newacronym{PL}{PL}{path loss}
\newacronym{AF}{AF}{array factor}
\newacronym{AH}{AH}{aerial highway}
\newacronym{PSO}{PSO}{particle swarm optimization}
\newacronym{eGA}{eGA}{elite genetic algorithm}
\newacronym{GA}{GA}{genetic algorithm}
\newacronym{ICC}{ICC}{international conference on communications}
\newacronym{SSB}{SSB}{synchronization signal block}
\newacronym{ES}{ES}{Eigenscore}
\newacronym{AoA}{AoA}{angle of arrival}
\newacronym{AoD}{AoD}{angle of departure}
\newacronym{UAV}{UAV}{unmanned aerial vehicle}
\newacronym{CCUAV}{CCUAV}{cellular connected unmanned aerial vehicle}
\newacronym{D2D}{D2D}{device to device}
\newacronym{gUE}{gUE}{ground user equipment}
\newacronym{UE}{UE}{user equipment}
\newacronym{MIMO}{MIMO}{multiple-input multiple-output}
\newacronym{mMIMO}{mMIMO}{massive multiple-input multiple-output}
\newacronym{MU-mMIMO}{MU-mMIMO}{multi-user massive multiple-input multiple-output}
\newacronym{MU-MIMO}{MU-MIMO}{multi-user multiple-input multiple-output}
\newacronym{SU-MIMO}{SU-MIMO}{single-user multiple-input multiple-output}
\newacronym{PRB}{PRB}{physical resource block}
\newacronym{RE}{RE}{resource element}
\newacronym{RSRP}{RSRP}{reference signal received power}
\newacronym{RSS}{RSS}{received signal strength}
\newacronym{mmWave}{mmWave}{millimetre wave}
\newacronym{eICIC}{eICIC}{enhanced inter-cell interference coordination}
\newacronym{SINR}{SINR}{signal-to-interference-plus-noise ratio}
\newacronym{UAM}{UAM}{urban air mobility}
\newacronym{QoS}{QoS}{quality of service}
\newacronym{LoS}{LoS}{line of sight}
\newacronym{NLoS}{NLoS}{not line of sight}
\newacronym{BVLoS}{BVLoS}{beyond visual line of sight}
\newacronym{DoF}{DoF}{degree of freedom}
\newacronym{ZF}{ZF}{zero forcing}
\newacronym{CSI}{CSI}{channel state information}
\newacronym{3GPP}{3GPP}{3rd Generation Partnership Project}
\newacronym{SVD}{SVD}{single value decomposition}
\newacronym{PBCH}{PBCH}{physical broadcast channel}
\newacronym{PDSCH}{PDSCH}{physical data shared channel}
\newacronym{thp}{thp}{throughput}
\newacronym{UMi}{UMi}{urban micro}
\newacronym{UMa}{UMa}{urban macro}
\newacronym{CAGR}{CAGR}{compound annual growth rate}
\newacronym{HO}{HO}{handover}
\newacronym{MNO}{MNO}{mobile network operator}
\newacronym{NOMA}{NOMA}{non-orthogonal multiple access}
\newacronym{BO}{BO}{Bayesian optimization}
\newacronym{ML}{ML}{machine learning}
\newacronym{FR1}{FR1}{frequency range 1}
\newacronym{SO}{SO}{southern}
\newacronym{E}{E}{eastern}
\newacronym{NE}{NE}{northeastern}
\newacronym{RAN}{RAN}{radio access network}
\newacronym{BS}{BS}{base station}
\newacronym{ISD}{ISD}{inter-site distance}
\newacronym{IGD}{IGD}{inter-grid distance}
\newacronym{IHD}{IHD}{inter-highway distance}
\newacronym{IUD}{IUD}{inter-UAV distance}
\newacronym{RRC}{RRC}{radio resource control}
\newacronym{PSS}{PSS}{primary synchronization signal}
\newacronym{SSS}{SSS}{secondary synchronization signal}
\newacronym{MINP}{MINP}{mixed-integer nonlinear problem}
\newacronym{PAHSS}{PAHSS}{Particle Aerial Highway Swarm Segmentation}
\newacronym{MAMA}{\texttt{MAMA}}{mMIMO-Aerial-Metric-Association}
\newacronym{URD}{URD}{urban random distributed}
\newacronym{UDN}{UDN}{ultra dense network}
\newacronym{ADAM}{ADAM}{ADAptive Moment estimation}
\newacronym{CRS}{CRS}{common reference signal}
\newacronym{MRT}{MRT}{maximum ratio transmission}
\newacronym{rhs}{rhs}{right hand side}
\newacronym{lhs}{lhs}{left hand side}
\newacronym{IoT}{IoT}{internet of thing}
\newacronym{KPI}{KPI}{key performance indicator}
\newacronym{NTN}{NTN}{non-terrestrial network}
\newacronym{TN}{TN}{terrestrial network}
\newacronym{HAPS}{HAPS}{high-altitude platform station}
\newacronym{CF-mMIMO}{CF-mMIMO}{cell free mMIMO}
\newacronym{EASA}{EASA}{European Union Aviation Safety Agency}
\newacronym{C2}{C2}{Command and Control}
\newacronym{UTM}{UTM}{unmanned aircraft system traffic management}

\newacronym{EE}{EE}{Energy Efficient}
\newacronym{SE}{SE}{Spectral Efficient}
\newacronym{UPV}{UPV}{Universitat Politecnica de Valencia}
\newacronym{RS}{RS}{reference signal}
\newacronym{PCI}{PCI}{physical cell identity}
\newacronym{RTT}{RTT}{round trip time}

\newacronym{SORA}{SORA}{Specific Operations Risk Assessment}

\newacronym{DAA}{DAA}{Detect and Avoid}
\newacronym{USSP}{USSP}{U-Space Service Provider}

\newacronym{SLA}{SLA}{service level agreement}

\newacronym{QoM}{QoM}{quality of mission}
\newacronym{QoE}{QoE}{quality of experience}
\newacronym{XR}{XR}{extended reality}

\newacronym{ISAC}{ISAC}{Integrated sensing and communication}

\newacronym{DT}{DT}{digital twin}
\newacronym{RRM}{RRM}{radio resource management}

\newacronym{ICT}{ICT}{information and communication technology}
\newacronym{RU}{RU}{radio unit}
\newacronym[plural=RATs,
            longplural=radio access technologies]
            {RAT}{RAT}{radio access technology}

\newacronym{BBU}{BBU}{baseband unit}
\newacronym{MCPA}{MCPA}{multi-carrier power amplifier}

\newacronym{CS}{CS}{carrier shutdown}

\newacronym{RL}{RL}{reinforcement learning}
\newacronym{LEO}{LEO}{low Earth orbit}

\newacronym{NGMN}{NGMN}{Next Generation Mobile Networks Alliance}

\newacronym{GP}{GP}{Gaussian Process}
\newacronym{TuRBO}{TuRBO}{trust region Bayesian optimization}
\newacronym{HDBO}{HDBO}{high-dimensional Bayesian optimization}

\newacronym{LogEI}{LogEI}{log expected improvement}

\setglossarystyle{list} 

\usepackage[normalem]{ulem}

\usepackage[caption=false,font=small,labelfont=sf,textfont=sf]{subfig} 
\usepackage{float}
\usepackage{tikz}
\usepackage{pgfplots}
\usetikzlibrary{patterns}
\usetikzlibrary{positioning,fit}
\pgfplotsset{compat=newest}

\usepackage[short]{optidef}

\usepackage{multirow}

\definecolor{ieeeblue}{RGB}{31,78,121}

\def\BibTeX{{\rm B\kern-.05em{\sc i\kern-.025em b}\kern-.08em
    T\kern-.1667em\lower.7ex\hbox{E}\kern-.125emX}}

\DeclareMathOperator*{\argmax}{\arg\!\max}

\newtheorem{Problem}{\bf Problem}

\renewcommand{\figurename}{Figure}

\pgfplotsset{compat=1.18}
\definecolor{darkslategray828178}{RGB}{82,81,78}
\definecolor{darkgray176}{RGB}{176,176,176}
\definecolor{gainsboro225224217}{RGB}{225,224,217}
\definecolor{lightgray204}{RGB}{204,204,204}
\definecolor{steelblue31119180}{RGB}{31,119,180}
\definecolor{indianred2097391}{RGB}{209,73,91}
\definecolor{turboorange}{RGB}{230,126,34}
\definecolor{communitygreen}{RGB}{39,139,91}
\colorlet{combinedtpzerodark}{steelblue31119180!82!black}
\colorlet{combinedtponedark}{turboorange!82!black}

\newlength\fwidth
\newlength\fheight

\pgfplotsset{
  ieeeaxis/.style={
    width=\fwidth,
    height=\fheight,
    tick align=outside,
    tick pos=left,
    axis background/.style={fill=white},
    axis line style={draw=black,line width=0.35pt},
    tick style={color=black,line width=0.35pt},
    xmajorgrids=true,
    ymajorgrids=true,
    x grid style={draw=darkgray176!50!white,dashed},
    y grid style={draw=darkgray176!50!white,dashed},
    xmin=-0.65,
    xmax=9.65,
    xtick={0,1,2,3,4,5,6,7,8,9},
    xticklabels={
      {H \\ (0.1,0.6)},
      {H \\ (0.3,0.7)},
      {H \\  (0.5,0.9)},
      {NH \\ (0.1,0.6)},
      {NH \\ (0.3,0.7)},
      {NH \\ (0.5,0.9)},
      {All Off},
      {No CS},
      {BO},
      {ComBO}
    },
    xticklabel style={font=\tiny, align=center,rotate=0},
    yticklabel style={font=\footnotesize},
    label style={font=\scriptsize},
    enlarge x limits=false,
    clip=false,
  },
  h6bar/.style={
    ybar,
    mark=none,
    bar width=7pt,
    bar shift=-4pt,
    draw=darkslategray828178,
    line width=0.35pt,
    fill=steelblue31119180,
  },
  h21bar/.style={
    ybar,
    mark=none,
    bar width=7pt,
    bar shift=4pt,
    draw=darkslategray828178,
    line width=0.35pt,
    fill=indianred2097391,
    postaction={pattern=north east lines,pattern color=darkslategray828178},
  },
  turboh6bar/.style={h6bar,fill=turboorange},
  turboh21bar/.style={h21bar,fill=turboorange},
  communityh6bar/.style={h6bar,fill=communitygreen},
  communityh21bar/.style={h21bar,fill=communitygreen},
  meanh6bar/.style={h6bar,fill=steelblue31119180!30!white},
  meanh21bar/.style={h21bar,fill=indianred2097391!30!white},
  meanturboh6bar/.style={h6bar,fill=turboorange!30!white},
  meanturboh21bar/.style={h21bar,fill=turboorange!30!white},
  meancommunityh6bar/.style={h6bar,fill=communitygreen!30!white},
  meancommunityh21bar/.style={h21bar,fill=communitygreen!30!white},
  combinedtpzeroh6bar/.style={h6bar,bar width=4pt,bar shift=-6.6pt,fill=combinedtpzerodark},
  combinedtpzeroh21bar/.style={h21bar,bar width=4pt,bar shift=-2.2pt,fill=combinedtpzerodark},
  combinedtponeh6bar/.style={h6bar,bar width=4pt,bar shift=2.2pt,fill=combinedtponedark},
  combinedtponeh21bar/.style={h21bar,bar width=4pt,bar shift=6.6pt,fill=combinedtponedark},
  sourceerrors/.style={
    error bars/.cd,
      y dir=both,
      y explicit,
      error bar style={draw=darkslategray828178,line width=0.5pt},
      error mark options={rotate=90,mark size=2.1pt,line width=0.5pt,draw=darkslategray828178},
  },
  p5errors/.style={
    error bars/.cd,
      y dir=minus,
      y explicit,
      error bar style={draw=darkslategray828178,line width=0.55pt},
      error mark options={rotate=90,mark size=2.2pt,line width=0.55pt,draw=darkslategray828178},
  },
}

\begin{document}

\title{ComBO: Community Bayesian Optimization for Carrier Shutdown in HAPS-Hypercell Networks}

\author{
\IEEEauthorblockN{
Matteo Bernabè\IEEEauthorrefmark{1}, 
Leonardo Spampinato\IEEEauthorrefmark{1}, 
David López-Pérez\IEEEauthorrefmark{1}
and 
Nicola Piovesan\IEEEauthorrefmark{7}
}
\normalsize\IEEEauthorblockA{\emph{
\IEEEauthorrefmark{1}Universitat Politècnica de València (UPV), Spain}, \quad \emph{
\IEEEauthorrefmark{7}Huawei Technologies, France} 
}

\thanks{This research is supported by the Generalitat Valenciana, Spain, through the CIDEGENT PlaGenT, Grant CIDEXG/2022/17, Project iTENTE, and the action CNS2023-144333, financed by MCIN/AEI/10.13039/501100011033 and the European Union “NextGenerationEU”/PRTR.} 
\vspace{-1em}
}

\maketitle

\begin{abstract}
    In this paper, we investigate \ac{CS} in \ac{HAPS}-Hypercell-assisted networks, leveraging network-wide information to optimize both coverage macro cells and capacity cells \ac{CS}.
    To address the high dimensionality of this problem, we propose \textit{ComBO}, a community-aware Bayesian optimization framework that exploits \ac{UE} \ac{RSRP} reports to construct a fallback graph representing the underlying network and identify groups of strongly coupled cells.
    The resulting communities decompose the original optimization problem into lower-dimensional subproblems, which are sequentially optimized using \ac{TuRBO} and refined through a final joint search. Evaluation in a high-fidelity 3GPP-based system-level environment shows that \textit{ComBO} consistently outperforms legacy \ac{CS} strategies and overcomes the scalability limitations of global \ac{BO}. 
    Considering normalized terrestrial power \textit{ComBO} achieves nearly 84\% savings while preserving quality of service.
    These results demonstrate the potential of community-aware decomposition as a scalable approach to large-scale network optimization.
\end{abstract}

\glsresetall

\section{Introduction}\label{sec:Intro}

The increasing density and traffic demand of modern cellular networks have made energy efficiency a key objective in \ac{RAN} design and management. 
As \ac{RAN} components account for a significant share of overall mobile network energy consumption,
efficient radio resource management while maintaining the required \ac{QoS} is essential for sustainable and cost-efficient mobile networks~\cite{GSMA_climateAction24}. 
Despite significant technological progress,  
the rate of improvement in global energy efficiency has slowed in recent years,
underscoring the need for more effective technologies, control strategies, and energy management approaches~\cite{IEA2026StateEnergyPolicy}. 
Consequently, energy-aware \ac{RAN} optimization has attracted increasing attention from \acp{MNO} and network manufacturers, 
with approaches ranging from radio resource allocation and base-station sleep modes to intelligent network control~\cite{9678321}. 
Developing optimization strategies that effectively balance energy consumption and network performance therefore remains a key research challenge and duty.
Among widely adopted energy-saving mechanisms~\cite{9678321},
\ac{CS} stands out as a major enabler: 
switching off the main radio components of a cell for extended periods substantially lowers consumption, particularly during low-traffic hours~\cite{9928089}.

As highlighted in our preliminary work~\cite{bernabe2026hapshypercellenablingcoverage},
a key limitation of conventional \ac{CS} lies in the always-on macro coverage cells, 
which must remain active to guarantee minimum coverage even when high-performing capacity cells are switched off during low traffic demand.
To overcome this constraint, we proposed the \ac{NTN} \ac{HAPS}-Hypercell concept. The Hypercell maintains continuous minimum coverage, allowing terrestrial macro cells to be switched off too and enabling more energy-efficient operation~\cite{bernabe2026hapshypercellenablingcoverage}.
However, our previous architecture exploited the benefits of the Hypercell while relying on decentralized decision-making based on limited local information, precluding a holistic, network-wide view~\cite{bernabe2026hapshypercellenablingcoverage}.
%
%
Centralized coordination would overcome this but at the cost of a high-dimensional problem: for reference,
in \ac{3GPP} deployments, in a few square kilometres, there may be hundreds of 4G and 5G cells, and thus as many optimization variables, if not more.
Such dimensionality poses a significant challenge for conventional optimization approaches, 
which quickly become intractable.

We therefore propose a novel framework that combines \ac{BO} with graph-based community detection to reduce problem dimensionality and enable efficient exploration of high-quality solutions in large-scale \ac{RAN} scenarios.
The main contributions are summarized as follows:
\begin{itemize}
    \item We extend the \ac{HAPS}-Hypercell~\cite{bernabe2026hapshypercellenablingcoverage} toward a centralized \ac{CS} control algorithm,
    where the Hypercell relays network information to a ground controller
    that jointly optimizes coverage and capacity carrier shutdown states.

    \item We propose \textit{ComBO}, 
    a community-aware \ac{BO} framework that exploits \ac{UE} \ac{RSRP} reports to construct a graph representation of the network and identify groups of strongly coupled cells. 
    The resulting communities decompose the original high-dimensional problem into smaller subproblems that are sequentially optimized.

    \item We evaluate the proposed framework in a high-fidelity,
    \ac{3GPP}-based system-level environment, 
    showing that community decomposition improves the energy consumption-\ac{QoS} trade-off over both legacy \ac{CS} strategies and global optimization.
\end{itemize}

\section{Hypercell Carrier Shutdown Framework}\label{sec:Hypercell_opt_framework}

\Ac{CS} is an energy-saving mechanism that temporarily deactivates underutilized carriers while maintaining service continuity through the remaining active layers. 
In this work, each cell operates on a single carrier through a dedicated \ac{RU}; 
therefore, applying \ac{CS} transitions the corresponding \ac{RU} to a shutdown state. 
In this state, only a subset of components remains powered, 
reducing the static power consumption associated with the cell.
Since static power accounts for a significant share of total consumption, 
\ac{CS} can substantially reduce overall network consumption~\cite{9928089,VALLERO2025111070}.

\subsection{Coverage-Capacity Layer and Hypercell Role}
Cellular networks typically comprise multiple cells operating across different \acp{RAT} and frequency bands,
which can be organized into two main layers:
capacity and coverage.
The \emph{capacity layer} consists of high-frequency, high-capacity cells deployed to serve localized demand in hotspot areas;
typically co-located with macro cells,
these cells operate well below capacity during off-peak periods and are therefore natural \ac{CS} candidates.
The \emph{coverage layer} consists of always-on macro cells at lower frequencies,
providing wide-area coverage and guaranteeing a minimum service level when capacity cells are deactivated.
The proposed \ac{HAPS}-Hypercell adds a third \ac{NTN} layer above the coverage one.
By sustaining the minimum required coverage over the whole area, 
it relieves macro cells of their always-on role,
allowing both coverage and capacity cells to be jointly switched off during low-demand periods~\cite{bernabe2026hapshypercellenablingcoverage}.



\subsection{Proposed Hypercell Framework}
As networks continuously move toward data-driven operation, 
cells are expected to build more and more \ac{ML} models from large volumes of network data,
including power consumption, load conditions, traffic patterns, and \ac{UE} behaviour. 
Rather than remaining local, 
these models are envisioned to be shared among network elements to enable continuous refinement, network-wide awareness, and adaptive energy management~\cite{3gppStudyArtificialIntelligence2023, 11590865}.
Building on this, 
we propose an architecture in which terrestrial cells paired with the \ac{HAPS}-Hypercell continuously collect measurements, train local models, and share information with the Hypercell, which aggregates these inputs and shares them with a central controller hosted in the ground network infrastructure.
Thus, the controller, combining these aggregated models and reports, constructs and continuously updates a network \ac{DT}.
Leveraging the shared models, the \ac{DT} can forecast network load states at a target hour, enabling the corresponding \ac{CS} configuration to be optimized offline, validated in virtual sandboxes, and deployed in advance.


\section{Community-aware Bayesian Optimization}\label{sec:CommunityAware_Bayesian_Optimization}

The main advantage of the proposed framework is the global observability of the system, 
which enables a centralized optimization pipeline and, in turn, the computation of global solutions. 
Specifically, the main optimization problem can be formulated as follows,
\begin{Problem}\label{opt:CS_Optimization}
Terrestrial Cell \ac{CS} Status Optimization
\begin{mini!}[2]
  {\mathbf{x}}
  {\displaystyle
   \mathbb{E}
   \left[
   \sum_{c \in \mathcal{C}_{\mathrm{tn}}}
   p_c\!\left(\mathbf{x},\boldsymbol{\xi}(\mathbf{x})\right)
   \right]}
  {}{}
  \addConstraint{
    \left|
    \mathbb{E}
    \left[
    R_u\!\left(\mathbf{x},\boldsymbol{\xi}(\mathbf{x})\right)
    \right]
    - R_u^{\star}
    \right|}
    {\leq \epsilon,}
    {\quad \forall u \in \mathcal{U} \label{C1.1}}
  \addConstraint{\mathbf{x}}
    {\in \{0,1\}^{\lvert \mathcal{C}_{\mathrm{tn}}\rvert}}
    {\label{C1.2}}
\end{mini!}
\end{Problem}
where $\mathbf{x}$ is the binary vector encoding the \acs{CS} status (i.e., on/off) of each terrestrial cell;
$\boldsymbol{\xi}(\mathbf{x})$ denotes the stochastic network state induced by the selected \ac{CS} configuration $\mathbf{x}$;
$p_c\!\left(\mathbf{x},\boldsymbol{\xi}(\mathbf{x})\right)$ denotes the power used by terrestrial cell $c$ under the resulting network state;
$\mathcal{U}$ and $\mathcal{C}_{\mathrm{tn}}$ denote the sets of \acp{UE} and terrestrial cells, respectively;
$R_u\!\left(\mathbf{x},\boldsymbol{\xi}(\mathbf{x})\right)$ and $R_u^{\star}$ denote the rate achieved by \ac{UE} $u$ and its demanded rate, respectively;
and $\epsilon > 0$ is the tolerance on the rate mismatch.

However, centralized approaches are known to scale poorly, 
particularly when solving non-trivial, complex, stochastic, and nonlinear optimization problems such as Problem~\ref{opt:CS_Optimization}.

\subsection{Bayesian Optimization and TuRBO}

Given the complexity of Problem~\ref{opt:CS_Optimization} and the high cost of evaluating candidate solutions, 
analytical or classical numerical methods are impractical.
\Ac{BO}, a class of surrogate-based heuristics for black-box optimization, is instead a natural candidate, 
as it is designed to efficiently search solutions when objective evaluations are expensive~\cite{7352306, garnett2023bayesian}.

\ac{BO} relies on Bayes theorem to conduct an informed search of the solution space.
It iteratively builds and updates a probabilistic approximation of the objective,
the \textit{surrogate model},
which is cheap to evaluate and can return both a score and its uncertainty for every point of the search space. 
Then, together with an \textit{acquisition function} that maps the surrogate posterior to a scalar score, 
it selects the next point of the search space to evaluate, 
thereby governing the exploration and exploitation trade-off.

However, the main limitation of \ac{BO} is the \textit{curse of dimensionality}~\cite{bellman1957dynamic}:
beyond a few tens of variables, 
its sample complexity grows rapidly and performance deteriorates~\cite{Binois2021A, SantoniML2024}.
Several high-dimensional \ac{BO} strategies address this, among which \ac{TuRBO}~\cite{NEURIPS2019_6c990b7a} emerges as the most promising across a wide family of problems,
both in terms of convergence and computation time.
Rather than fitting a single global surrogate, 
\ac{TuRBO} maintains multiple parallel local runs,
each with an independent model confined to a \textit{trust region} centered on a promising point.
Trust regions expand after successful evaluations and shrink after unsuccessful ones, 
with candidates selected by optimizing an acquisition function within each region.

Nevertheless, above approximately 60 variables,
\ac{TuRBO} also begins to suffer the inevitable \textit{curse of dimensionality}~\cite{SantoniML2024}.
Since a few square kilometres of urban deployment already contain hundreds of cells,
directly solving Problem~\ref{opt:CS_Optimization} is challenging in both computational budget and solution quality.
We therefore propose \textit{ComBO}, 
a heuristic pipeline that leverages the aggregated reports to construct a graph representation of the network and identify areas over which smaller, dimension-limited \ac{TuRBO} searches can be run.

\begin{figure*}[!t]
    \centering
    \resizebox{\textwidth}{!}{%
        \begin{tikzpicture}[
    node distance=1.2cm,
    auto,
    >=stealth
]


\node[
    draw=blue!55!black,
    thick,
    rounded corners=2pt,
    fill=blue!8,
    minimum width=2cm,
    minimum height=3.5cm,
    text width=2cm,
    align=center
] (A) {\texttt{Giulia} \\[2pt] DT};


\node[
    draw=blue!55!black,
    thick,
    rounded corners=2pt,
    fill=blue!8,
    minimum width=2.5cm,
    minimum height=1.2cm,
    text width=2.5cm,
    align=center,
    right=0.6cm of A
] (B) {
    \textbf{Leiden Community Detection}\\[3pt]
    $\mathcal{C}_{\mathrm{tn}} = \bigcup_{k=1}^{K} \mathcal{C}_k$
};


\node[
    draw=orange!65!black,
    very thick,
    rounded corners=6pt,
    fill=orange!5,
    minimum width=5cm,
    minimum height=3.5cm,
    right=1.5cm of B
] (C) {};

\node[
    text=orange!70!black,
    font=\small\bfseries
] at ([yshift=1.35cm]C.center)
{Block-wise Optimization};


\node[
    draw=orange!65!black,
    thick,
    rounded corners=2pt,
    fill=orange!10,
    minimum width=3cm,
    minimum height=0.9cm,
    align=center
] (C1) at ([yshift=0.75cm]C.center)
{\textbf{TuRBO}};


\node[
    draw=orange!65!black,
    thick,
    rounded corners=2pt,
    fill=orange!10,
    minimum width=3cm,
    minimum height=1.1cm,
    text width=3cm,
    align=center
] (C2) at ([yshift=-0.75cm]C.center)
{
    \textbf{Community}\\[2pt]
    $(\mathcal{C}_k,\mathbf{x}_k)$
};


\node[
    draw=green!50!black,
    thick,
    rounded corners=2pt,
    fill=green!10,
    minimum width=2.7cm,
    minimum height=1.2cm,
    text width=2.7cm,
    align=center,
    right=1.5cm of C
] (D) {
    \textbf{Joint Polish TuRBO}\\[3pt]
    $(\mathcal{C}_{\mathrm{tn}},\mathbf{\hat{x}})$
};


\node[
    draw=green!50!black,
    thick,
    rounded corners=2pt,
    fill=green!10,
    minimum width=2.7cm,
    minimum height=1.2cm,
    text width=2.7cm,
    align=center,
    right=1.5cm of D
] (E) {
    \textbf{Final Solution}\\[3pt]
    $(\mathcal{C}_{\mathrm{tn}},\mathbf{x}^*)$
};


\draw[
    ->,
    very thick,
    draw=black!65
] (A) -- (B);

\draw[
    ->,
    very thick,
    draw=black!65
] (B) --
    node[midway, above]
    {$\{\mathcal{C}_k\}_{k=1}^{K}$}
    (C.west);

\draw[
    ->,
    very thick,
    draw=black!65
] (C.east) --
    node[midway, above]
    {$(\mathcal{C}_{\mathrm{tn}},\mathbf{\hat{x}})$}
    (D);

\draw[
    ->,
    very thick,
    draw=black!65
] (D) --
    node[midway, above]
    {$(\mathcal{C}_{\mathrm{tn}},\mathbf{x}^*)$}
    (E);


\draw[
    ->,
    thick,
    draw=orange!70!black
] ([xshift=-0.15cm]C1.south) --
  ([xshift=-0.15cm]C2.north);

\draw[
    ->,
    thick,
    draw=orange!70!black
] ([xshift=0.15cm]C2.north) --
  ([xshift=0.15cm]C1.south);


\node[
    draw=black!35,
    very thick,
    rounded corners=8pt,
    fill=none,
    inner sep=0.55cm,
    fit=(A) (B) (C) (D) (E),
    label={
        [yshift=-0.05cm]
        above:{\large\bfseries  \textit{ComBO} - Community Bayesian Optimization}
    }
] (outer) {};

\end{tikzpicture}
    }
    \caption{\textit{ComBO} pipeline: the fallback graph derived from measurements is partitioned into communities, optimized sequentially, and finally refined.}
    \label{fig:pipeline_turbo_community}
    \vspace{-1.5em}
\end{figure*}
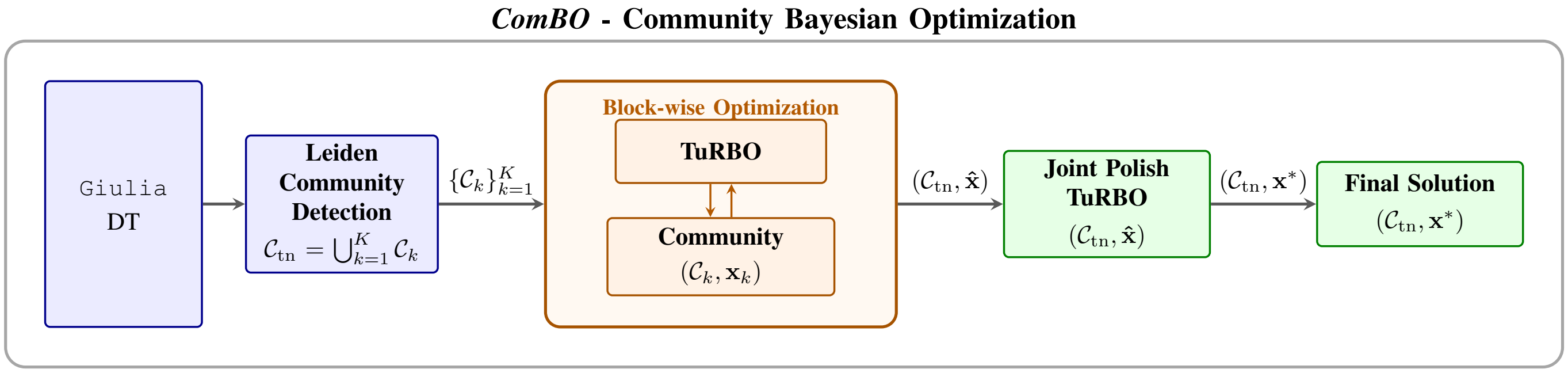

\subsection{Network Graph and Community Detection}

To reduce the dimensionality of Problem~\ref{opt:CS_Optimization} and enable the effective application of solvers such as \ac{TuRBO},
we adopt an architecture in which each terrestrial cell is paired with the \ac{HAPS}-Hypercell and with which it can exchange information.

Specifically, we assume that each terrestrial cell collects beam-level \ac{RSRP} reports from its connected \acp{UE}.
We build on the beam-management framework, 
in which each \ac{UE} measures the Layer-1 \ac{RSRP} on its configured \ac{RS} resources, 
as defined in~\cite{3GPP38215}, 
and reports the measurements to the network. 
Each cell forwards the collected reports through the Hypercell to the controller, which aggregates them over time to update the \ac{DT} with a comprehensive representation of the \ac{RSRP} measurements reported by \acp{UE}.
Layer-1 \ac{RSRP} measurements are not limited to the serving cell, 
but may also include measurements from neighboring cells. 
By aggregating these reports, 
the \ac{DT} and controller can therefore construct a representation of the local radio environment and the interactions among cells.
Since \acp{UE} typically report only measurements associated with detectable cells, 
\ac{RSRP} values from sufficiently distant cells are generally absent, 
likely due to their negligible received power.

\subsubsection{Fallback Graph}
From the collected \ac{RSRP} measurements,
we derive a graph representation of the network based on the fallback relationships among cells, 
computed independently for each coverage and capacity layer.
For each cell $c$, we first identify the \acp{UE} associated with it according to the maximum-\ac{RSRP} criterion. We then consider cell $c$ unavailable and re-associate these \acp{UE} to the cell $c'$ in layer $l$ providing the strongest \ac{RSRP}.
Thus, the resulting fallback weight $w_{c,c'}^{(l)}$ is
\begin{equation}
w_{c,c'}^{(l)} = \frac{\left|\hat{\mathcal{U}}_{c'}^{(l)}\right|}
{\left|\mathcal{U}_{c}\right|},
\end{equation}
where $\left|\mathcal{U}_{c}\right|$ denotes the cardinality of the set of \acp{UE} originally associated with cell $c$, 
and $\left|\hat{\mathcal{U}}_{c'}^{(l)}\right|$ the cardinality of the subset re-associated with cell $c'$ on layer $l$ after removing cell $c$.
The weight thus represents the fraction of \acp{UE} of cell $c$ falling back to cell $c'$ on layer $l$.

The layer-specific fallback weights are then concatenated to form the fallback vector associated with cell $c$:
\begin{equation}
\mathbf{f}_{c}
=
\left[ 
    \mathbf{w}_{c}^{(1)} \;\middle|\; \mathbf{w}_{c}^{(2)} \;\middle|\; \cdots \;\middle|\; \mathbf{w}_{c}^{(L)}
\right],
\end{equation}
in which $\mathbf{w}_{c}^{(l)} = [w_{c,1}^{(l)},\ldots,w_{c,N^{(l)}}^{(l)}]$ 
and 
$N^{(l)}$ is the number of cells in layer $l$.
Finally, stacking the fallback vectors of all $|\mathcal{C}_{\mathrm{tn}}|$ cells yields the fallback matrix
\begin{equation} \label{eq:fallback_adjacency_matrix}
\mathbf{F} = 
\begin{bmatrix}
\mathbf{f}_{1}, \mathbf{f}_{2}, \ldots, \mathbf{f}_{|\mathcal{C}_{\mathrm{tn}}|} \end{bmatrix}^{\rm T},
\end{equation}
which is used as the weighted adjacency representation of the fallback graph.
It should be noted that the resulting adjacency matrix $\mathbf{F}$ is square but not necessarily symmetric,
since the fallback relationship between two cells depends on the propagation conditions experienced by their respective associated \acp{UE}, 
which are generally asymmetric.

\subsubsection{Graph Communities}

Community detection partitions the vertex set $\mathcal{V}$ of a graph $\mathcal{G}=(\mathcal{V},\mathcal{E})$ into groups of nodes exhibiting stronger interactions within the same group than with nodes of other groups~\cite{fortunato2010community}.
Here, $\mathcal{V}$ comprises all network cells, and
the fallback matrix $\mathbf{F}$ defined by
eq.~\eqref{eq:fallback_adjacency_matrix} is the weighted adjacency
matrix of the graph, whose directed edges represent fallback relations.
On this graph,
a community is a group of cells with strong mutual fallback:
when one of them becomes unavailable,
its \acp{UE} are likely to be re-associated with cells of the same community.
Grouping such strongly coupled cells therefore allows their
carrier-shutdown decisions to be jointly optimized while reducing
the dimensionality of the original problem.
Communities are identified with the Leiden algorithm~\cite{Traag2019},
applied directly to the weighted directed graph.
We employ the Reichardt--Bornholdt configuration model,
whose quality function for a directed graph and a partition $\mathcal{P}$ of the vertex set $\mathcal{V}$ is
\begin{equation}
Q(\mathcal{P}) =
\sum_{i=1}^{|\mathcal{V}|}
\sum_{j=1}^{|\mathcal{V}|}
\left(
F_{ij}
-
\gamma
\frac{k_i^{\mathrm{out}} k_j^{\mathrm{in}}}{m}
\right)
\delta(g_i,g_j),
\label{eq:quality_function}
\end{equation}
where
\begin{equation}
k_i^{\mathrm{out}}
=
\sum_{j=1}^{|\mathcal{V}|} F_{ij},
\quad
k_i^{\mathrm{in}}
=
\sum_{j=1}^{|\mathcal{V}|} F_{ji},
\quad {\rm and} \quad
m
=
\sum_{i=1}^{|\mathcal{V}|}
\sum_{j=1}^{|\mathcal{V}|}
F_{ij}
\end{equation}
are the weighted outgoing and incoming strengths of node $i$ and the total graph weight, $g_i$ is the community assigned to cell $i$, and $\delta(g_i,g_j)=1$ if $g_i=g_j$ and zero otherwise. 
The resolution parameter $\gamma$ controls the granularity of the partition, 
with larger values favouring more and smaller communities.
Starting from an initial partition, 
Leiden iterates three phases: 
\textit{i)} local movement of nodes between communities, 
\textit{ii)} refinement of the resulting partition, and 
\textit{iii)} aggregation into a reduced graph. 
The iterative process is repeated until no further improvement of eq.~\eqref{eq:quality_function} is obtained~\cite{Traag2019}.

\subsection{ComBO - Community Bayesian Optimization}

Building on \ac{TuRBO} and the community detection, 
we propose \textit{ComBO},
a  framework that decomposes the original \ac{CS} problem into lower-dimensional community-level searches.
As illustrated in Fig.~\ref{fig:pipeline_turbo_community},
it consists of five major components:
\begin{enumerate}
    \item \textbf{\texttt{Giulia} DT:}
    The digital twin,
    where the reports collected from all cells are aggregated,
    instantiates the multi-layer network for the target condition and provides the environment in which candidate \ac{CS} configurations are evaluated. 
    The baseline \ac{CS} strategies of~\cite{bernabe2026hapshypercellenablingcoverage} are evaluated first here,
    so that \ac{TuRBO} is initialized from the best available configuration.
    
    \item \textbf{Leiden Community Detection:}
    The fallback graph is built and partitioned into $K$ communities, 
    processed sequentially starting from the largest.
    
    \item \textbf{Community-level \ac{TuRBO}:}
    A dedicated \ac{TuRBO} search is performed over the \ac{CS} states of the current community $\mathcal{C}_k$.
    As the variables are binary, 
    the trust region limits the Hamming distance between a candidate and the current network configuration. 
    Although only cells of $\mathcal{C}_k$ are modified, 
    each candidate is evaluated over the complete network, 
    yielding a network-level score comparable across subproblems.
    
    \item \textbf{Sequential Community Sweep:}
    After optimizing community $\mathcal{C}_k$, 
    the selected carrier-state changes are applied to the current network configuration before proceeding to the next community $\mathcal{C}_{k+1}$. 
    Once all $K$ communities have been processed,
    a complete \emph{community sweep} is concluded.
    If the improvement achieved during the sweep is below a predefined threshold, 
    a perturbed restart is performed by randomly modifying a small number of carrier states in the best configuration found so far;
    this is performed for a limited number of rounds,
    $N_{\rm rnd}^{\rm comm}$, set to two here.
    
    \item \textbf{Joint Polish \ac{TuRBO}:}
    A \ac{TuRBO} search over all $|\mathcal{C}_{\mathrm{tn}}|$ cells,
    started from the best configuration found in the preceding sweeps and confined to a tight Hamming trust region. 
    It allows simultaneous modification of cells in different communities, 
    thereby capturing cross-community interactions that no single community-level search can explore.
\end{enumerate}

\section{Simulation and Optimization Settings}\label{sec:sim_and_opt_settings}

In the following, 
we describe the simulation network settings, the objective function adopted to drive the \textit{ComBO} optimization process, and the optimization settings used for the Leiden community-detection algorithm and \ac{TuRBO}.

\subsection{Environment Network Model}

We rely on \texttt{Giulia},
an open-source system-level simulator that couples expert-driven, \ac{3GPP}-compliant, and AI-based models for heterogeneous cellular networks.\footnote{\texttt{Giulia} is publicly available at \url{https://github.com/giulia-open-lab/OpenGiuliaSLS.git}.} 
In this work, \texttt{Giulia} realizes the \ac{DT} of Section~\ref{sec:Hypercell_opt_framework}, 
providing the environment in which candidate \ac{CS} configurations are evaluated.
We adopt the network scenario of~\cite{bernabe2026hapshypercellenablingcoverage}: 
a two-tier \ac{3GPP} \ac{UMa} deployment of 19 tri-sectorized sites with 500\,m inter-site distance, 
hosting co-located 4G and 5G at 2 and 3.5\,GHz, for a total of $|\mathcal{C}_{\mathrm{tn}}| = 114$ terrestrial cells, 
with the 5G layer acting as capacity layer and the 4G layer as coverage layer. 
A single-sector \ac{HAPS} at the network centre,
equipped with a reflector antenna working at 1.8\,GHz, 
covers the entire underlying area.
The main difference with respect to~\cite{bernabe2026hapshypercellenablingcoverage} is that the \ac{HAPS} is here equipped with a reflector antenna deployed at 20\,km altitude,
with propagation models and parameters adjusted according to~\cite{3GPP38811}; 
we refer to~\cite{bernabe2026hapshypercellenablingcoverage} for the remaining deployment details, omitted here for space.

Then, to reflect the variability of commercial hardware,
the power model parameters of each \ac{RU} are independently randomized and kept fixed throughout optimization and simulation,
so that each cell exhibits a persistent power-consumption profile and thus a distinct power-saving potential under \ac{CS}.

To capture the temporal variability of real networks, 
the total number of \acp{UE} at a given hour follows the profile adopted in~\cite{bernabe2026hapshypercellenablingcoverage, EARTH_D23_2010}. 
To capture the spatial heterogeneity of real networks, 
those \acp{UE} are non-uniformly distributed over the area. 
Specifically, their locations are generated according to a continuous spatial density field with Gaussian-shaped hotspots,
resulting in areas with different concentrations.

\subsection{The Objective Function}

To solve Problem~\ref{opt:CS_Optimization},
we reformulate the original constrained problem through a scalar objective that jointly accounts for the terrestrial network consumption and the satisfaction of the served
\acp{UE}.
For each \ac{UE} $u\in\mathcal{U}$,
we define its satisfaction as
\begin{equation}
s_u(\mathbf{x}) =
\min\left(
\frac{R_u(\mathbf{x})}{R_u^{\star}},
1
\right).
\end{equation}
%


The network satisfaction, here representing network \ac{QoS}, is defined as the average satisfaction over all \acp{UE}, i.e.,
$S(\mathbf{x})=\frac{1}{|\mathcal{U}|}\sum_{u\in\mathcal{U}}s_u(\mathbf{x})$.
Similarly, the terrestrial network consumption is normalized with
respect to its range as
\begin{equation}\label{eq:terrestrial_norm_power}
\widetilde{P}(\mathbf{x}) =
\frac{
P_{\mathrm{tn}}(\mathbf{x})-P_{\rm min}
}{
P_{\rm max}-P_{\rm min}
}
\end{equation}
where $P_{\rm min}$ and $P_{\rm max}$ denote the network power consumption with all terrestrial cells shutdown and fully active under full load, respectively.

The final optimization objective is formulated as 

\begin{equation}\label{eq:ObjectiveFunction}
\mathbf{x}^{*}
=
\argmax_{\mathbf{x}\in\{0,1\}^{|\mathcal{C}_{\mathrm{tn}}|}}
\left[
S(\mathbf{x})
-
\widetilde{P}(\mathbf{x})
\right]\;,
\end{equation}
where $\widetilde{P}(\mathbf{x})$ promotes energy savings, while maximizing $S(\mathbf{x})$ preserves \ac{QoS}. 
Since both are normalized in $[0,1]$, 
they equally contribute to the optimization objective.

It should be noted that, for each community-level optimization, the optimized carrier states are embedded into the complete network configuration vector $\mathbf{x}$, while retaining the current states of all remaining cells. 
The objective function in eq.~\eqref{eq:ObjectiveFunction} is then evaluated on this full-network configuration,
ensuring that solutions obtained from different communities are assessed on a common network-wide basis.

\subsection{Optimization Settings}

For community detection, the Leiden algorithm is applied with
resolution parameter $\gamma=1$. For the considered
scenario~\cite{bernabe2026hapshypercellenablingcoverage}, this
results in seven communities containing between $6$ and $20$ cells
each. The main \textit{ComBO} optimization parameters are summarized in
Table~\ref{tab:optimization_settings}.
For each community, \ac{TuRBO} operates with $N_{\rm TR}$ trust
regions, whose Hamming radii are adapted after $N_{\rm succ}$
consecutive successful or $N_{\rm fail}$ unsuccessful evaluations.
Here, the \ac{LogEI} is adopted as the acquisition function.
The evaluation budget is divided into an initial and a model-guided
search phase, both of which scale with community size. Thus, larger
communities receive more evaluations while excessive sampling of
smaller ones is avoided.
After each community sweep, the objective improvement is evaluated.
When it falls below $\varepsilon_{\rm sweep}$, a perturbed restart
with $N_{\rm flip}$ carrier-state flips is performed. Finally, the
joint polish \ac{TuRBO} uses an initial Hamming radius
$r_{\rm init}^{\rm polish}$ to refine the best solution.

\begin{table}[!th]
\vspace{-1.5em}
\centering
\caption{\textnormal{\textit{ComBO}} optimization settings.}
\vspace{-1em}
\label{tab:optimization_settings}
\begin{tabular}{lclc}
\hline
\textbf{Parameter} & \textbf{Value} &
\textbf{Parameter} & \textbf{Value} \\
\hline
$\gamma$ & $1$ &
$N_{\rm TR}$ & $3$ \\

$(r_{\rm init},r_{\rm min},r_{\rm max})$ & $(3,1,8)$ &
$r_{\rm init}^{\rm polish}$ & $2$ \\

$(N_{\rm succ},N_{\rm fail})$ & $(3,10)$ &
$\Delta r$ & $1$ \\

Acquisition function & LogEI &
$\varepsilon_{\rm sweep}$ & $10^{-4}$ \\

$N_{\rm flip}$ & $5$ &
$N_{\rm rnd}^{\rm comm}$ & $2$ \\
\hline
\end{tabular}
\vspace{-1.5em}
\end{table}

\section{Results}\label{sec:Results}

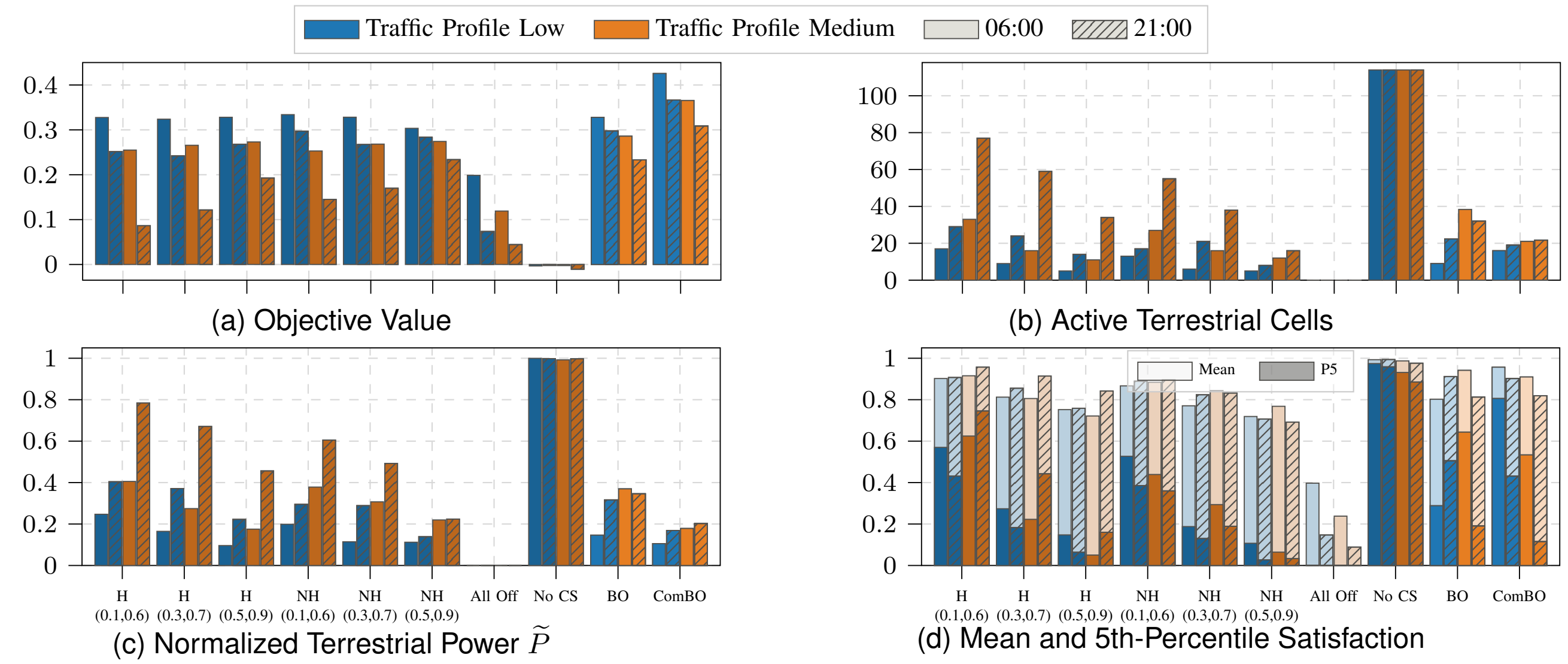
\begin{figure*}[!t]
  \centering
  \def\CombinedScenarioLegend{scenariolegendcombined}
  \def\CombinedSatisfactionLegend{1}

  \ref{scenariolegendcombined}\\[-0.75em]

  \subfloat[Objective Value\label{subfig:combined-reward}]{%
    \setlength\fwidth{0.975\columnwidth}%
    \setlength\fheight{0.45\columnwidth}%
    \makebox[\columnwidth][r]{\begin{tikzpicture}[trim axis left,trim axis right]
\ifdefined\CombinedSatisfactionLegend
  \def\CombinedLegendColumns{4}
\else
  \def\CombinedLegendColumns{6}
\fi
\begin{axis}[
  ieeeaxis,
  ymin=-0.035,
  ymax=0.45,
  ytick={0,0.1,...,0.4},
  xticklabels={},
  legend cell align={left},
  legend to name=\CombinedScenarioLegend,
  legend columns=\CombinedLegendColumns,
  legend style={fill=white,draw=lightgray204,font=\footnotesize,/tikz/every even column/.append style={column sep=0.7em}},
]
\addlegendimage{area legend,draw=darkslategray828178,line width=0.35pt,fill=steelblue31119180}
\addlegendentry{Traffic Profile Low}
\addlegendimage{area legend,draw=darkslategray828178,line width=0.35pt,fill=turboorange}
\addlegendentry{Traffic Profile Medium}
\addlegendimage{area legend,draw=darkslategray828178,line width=0.35pt,fill=gainsboro225224217}
\addlegendentry{06:00}
\addlegendimage{area legend,draw=darkslategray828178,line width=0.35pt,fill=gainsboro225224217,postaction={pattern=north east lines,pattern color=darkslategray828178}}
\addlegendentry{21:00}
\ifdefined\CombinedSatisfactionLegend\else
\addlegendimage{area legend,draw=darkslategray828178,line width=0.35pt,fill=darkslategray828178!25!white}
\addlegendentry{Mean satisfaction}
\addlegendimage{area legend,draw=darkslategray828178,line width=0.35pt,fill=darkslategray828178}
\addlegendentry{p5 satisfaction}
\fi

\addplot+[combinedtpzeroh6bar,forget plot] table[x=x,y=y]{
x y
0 0.327528031305
1 0.323789889784
2 0.328016587527
3 0.333890613551
4 0.328168546838
5 0.303378027092
6 0.198649340365
7 -0.003284829251
8 0.327853735453
9 0.425673776410
};
\addplot+[combinedtpzeroh21bar,forget plot] table[x=x,y=y]{
x y
0 0.251525525257
1 0.242039711533
2 0.267989624193
3 0.297110679675
4 0.267444202202
5 0.283918588012
6 0.073641386703
7 -0.002153619378
8 0.297561858286
9 0.366523941574
};
\addplot+[combinedtponeh6bar,forget plot] table[x=x,y=y]{
x y
0 0.254939376
1 0.265615411
2 0.273041783
3 0.252886849
4 0.268218807
5 0.274168364
6 0.119022799
7 -0.00245805845
8 0.285901582
9 0.365444543
};
\addplot+[combinedtponeh21bar,forget plot] table[x=x,y=y]{
x y
0 0.0865806586
1 0.12151001
2 0.192611524
3 0.144978881
4 0.169896729
5 0.233759076
6 0.0444084356
7 -0.0106742406
8 0.233209118
9 0.308449606
};
\addplot+[combinedtpzeroh6bar,fill=steelblue31119180,forget plot] coordinates {(8,0.327853735453) (9,0.425673776410)};
\addplot+[combinedtpzeroh21bar,fill=steelblue31119180,forget plot] coordinates {(8,0.297561858286) (9,0.366523941574)};
\addplot+[combinedtponeh6bar,fill=turboorange,forget plot] coordinates {(8,0.285901582) (9,0.365444543)};
\addplot+[combinedtponeh21bar,fill=turboorange,forget plot] coordinates {(8,0.233209118) (9,0.308449606)};
\end{axis}
\end{tikzpicture}}%
  }
  \hfill
  \subfloat[Active Terrestrial Cells\label{subfig:combined-active-cells}]{%
    \setlength\fwidth{0.975\columnwidth}%
    \setlength\fheight{0.45\columnwidth}%
    \makebox[\columnwidth][r]{\begin{tikzpicture}[trim axis left,trim axis right]
\begin{axis}[
  ieeeaxis,
  ymin=0,
  ymax=118,
  ytick={0,20,40,60,80,100,120},
  xticklabels={},
]
\addplot+[combinedtpzeroh6bar,forget plot] table[x=x,y=y]{
x y
0 17
1 9
2 5
3 13
4 6
5 5
6 0
7 114
8 9
9 16
};
\addplot+[combinedtpzeroh21bar,forget plot] table[x=x,y=y]{
x y
0 29
1 24
2 14
3 17
4 21
5 8
6 0
7 114
8 22.333333
9 19
};
\addplot+[combinedtponeh6bar,forget plot] table[x=x,y=y]{
x y
0 33
1 16
2 11
3 27
4 16
5 12
6 0
7 114
8 38.3333333
9 21
};
\addplot+[combinedtponeh21bar,forget plot] table[x=x,y=y]{
x y
0 77
1 59
2 34
3 55
4 38
5 16
6 0
7 114
8 32
9 21.6666667
};
\addplot+[combinedtpzeroh6bar,fill=steelblue31119180,forget plot] coordinates {(8,9) (9,16)};
\addplot+[combinedtpzeroh21bar,fill=steelblue31119180,forget plot] coordinates {(8,22.333333) (9,19)};
\addplot+[combinedtponeh6bar,fill=turboorange,forget plot] coordinates {(8,38.3333333) (9,21)};
\addplot+[combinedtponeh21bar,fill=turboorange,forget plot] coordinates {(8,32) (9,21.6666667)};
\end{axis}
\end{tikzpicture}}%
  }

  \par\vspace{-1em}
  \subfloat[Normalized Terrestrial Power $\widetilde{P}$\label{subfig:combined-normalized-power}]{%
    \setlength\fwidth{0.975\columnwidth}%
    \setlength\fheight{0.45\columnwidth}%
    \makebox[\columnwidth][r]{\begin{tikzpicture}
\ifdefined\HideCombinedMiddleXLabels
  \pgfplotsset{combinednormalizedaxis/.style={xticklabels={}}}
\else
  \pgfplotsset{combinednormalizedaxis/.style={}}
\fi
\begin{axis}[
  ieeeaxis,
  combinednormalizedaxis,
  ymin=0,
  ymax=1.05,
  ytick={0,0.2,0.4,0.6,0.8,1.0},
]
\addplot+[combinedtpzeroh6bar,forget plot] table[x=x,y=y]{
x y
0 0.247174728411
1 0.164728252700
2 0.096055488574
3 0.198923950788
4 0.114280820263
5 0.112213747370
6 0.000000000000
7 0.999652518516
8 0.146428568520
9 0.105445760522
};
\addplot+[combinedtpzeroh21bar,forget plot] table[x=x,y=y]{
x y
0 0.404271566603
1 0.371184472513
2 0.222663399469
3 0.295275867074
4 0.288980308133
5 0.138988960548
6 0.000000000000
7 0.998226579053
8 0.316614482886
9 0.168809213085
};
\addplot+[combinedtponeh6bar,forget plot] table[x=x,y=y]{
x y
0 0.405531161
1 0.274478241
2 0.175360975
3 0.378334206
4 0.30710798
5 0.219783143
6 0
7 0.992175714
8 0.369801796
9 0.178732152
};
\addplot+[combinedtponeh21bar,forget plot] table[x=x,y=y]{
x y
0 0.78396736
1 0.670940595
2 0.456438808
3 0.604995578
4 0.49221238
5 0.22399738
6 0
7 0.997370568
8 0.346014063
9 0.20228983
};
\addplot+[combinedtpzeroh6bar,fill=steelblue31119180,forget plot] coordinates {(8,0.146428568520) (9,0.105445760522)};
\addplot+[combinedtpzeroh21bar,fill=steelblue31119180,forget plot] coordinates {(8,0.316614482886) (9,0.168809213085)};
\addplot+[combinedtponeh6bar,fill=turboorange,forget plot] coordinates {(8,0.369801796) (9,0.178732152)};
\addplot+[combinedtponeh21bar,fill=turboorange,forget plot] coordinates {(8,0.346014063) (9,0.20228983)};
\end{axis}
\end{tikzpicture}}%
  }
  \hfill
  \subfloat[Mean and 5th-Percentile Satisfaction\label{subfig:combined-satisfaction}]{%
    \setlength\fwidth{0.975\columnwidth}%
    \setlength\fheight{0.45\columnwidth}%
    \makebox[\columnwidth][r]{\begin{tikzpicture}
\ifdefined\CombinedSatisfactionLegend
  \pgfplotsset{
    combinedsatisfactionlegend/.style={
      legend cell align={left},
      legend columns=2,
      legend style={
        at={(0.5,0.99)},
        anchor=north,
        fill=white,
        fill opacity=0.5,
        text opacity=1,
        draw=lightgray204,
        font=\tiny,
        /tikz/every even column/.append style={column sep=0.55em},
      },
    }
  }
\else
  \pgfplotsset{combinedsatisfactionlegend/.style={}}
\fi
\begin{axis}[
  ieeeaxis,
  combinedsatisfactionlegend,
  ymin=0,
  ymax=1.05,
  ytick={0,0.2,0.4,0.6,0.8,1.0},
]
\ifdefined\CombinedSatisfactionLegend
\addlegendimage{area legend,draw=darkslategray828178,line width=0.35pt,fill=darkslategray828178!25!white,
fill opacity=0.15}
\addlegendentry{Mean}
\addlegendimage{area legend,draw=darkslategray828178,line width=0.35pt,fill=darkslategray828178}
\addlegendentry{P5}
\fi
\addplot+[combinedtpzeroh6bar,fill=combinedtpzerodark!30!white,forget plot] table[x=x,y=y]{
x y
0 0.902230791020
1 0.812308032269
2 0.752088663629
3 0.866705177889
4 0.770617913938
5 0.718969801554
6 0.397298680731
7 0.993082860013
8 0.802136039426
9 0.956793313342
};
\addplot+[combinedtpzeroh21bar,fill=combinedtpzerodark!30!white,forget plot] table[x=x,y=y]{
x y
0 0.907322617117
1 0.855263895580
2 0.758642647855
3 0.889497226424
4 0.823868712538
5 0.706826136571
6 0.147282773406
7 0.993919340298
8 0.911738199458
9 0.901857096233
};
\addplot+[combinedtponeh6bar,fill=combinedtponedark!30!white,forget plot] table[x=x,y=y]{
x y
0 0.915409913
1 0.805709064
2 0.721444541
3 0.884107903
4 0.843545595
5 0.768119871
6 0.238045598
7 0.987259597
8 0.941604959
9 0.909621238
};
\addplot+[combinedtponeh21bar,fill=combinedtponedark!30!white,forget plot] table[x=x,y=y]{
x y
0 0.957128678
1 0.913960616
2 0.841661856
3 0.894953341
4 0.832005838
5 0.691515532
6 0.0888168712
7 0.976022087
8 0.812432299
9 0.819189042
};
\addplot+[combinedtpzeroh6bar,fill=steelblue31119180!30!white,forget plot] coordinates {(8,0.802136039426) (9,0.956793313342)};
\addplot+[combinedtpzeroh21bar,fill=steelblue31119180!30!white,forget plot] coordinates {(8,0.911738199458) (9,0.901857096233)};
\addplot+[combinedtponeh6bar,fill=turboorange!30!white,forget plot] coordinates {(8,0.941604959) (9,0.909621238)};
\addplot+[combinedtponeh21bar,fill=turboorange!30!white,forget plot] coordinates {(8,0.812432299) (9,0.819189042)};

\addplot+[combinedtpzeroh6bar,forget plot] table[x=x,y=y]{
x y
0 0.569871219062
1 0.273718542871
2 0.147598507143
3 0.526866584796
4 0.187712140535
5 0.107130449606
6 0.000138949749
7 0.974233301237
8 0.288356077783
9 0.805774657203
};
\addplot+[combinedtpzeroh21bar,forget plot] table[x=x,y=y]{
x y
0 0.431296898871
1 0.182811687987
2 0.063889165166
3 0.385388655274
4 0.130290621811
5 0.026918135717
6 0.000000000000
7 0.958570200743
8 0.505048441824
9 0.431131731383
};
\addplot+[combinedtponeh6bar,forget plot] table[x=x,y=y]{
x y
0 0.624653439
1 0.222708644
2 0.050327296
3 0.439122613
4 0.293873701
5 0.064353675
6 0.000080551
7 0.9313695545
8 0.643482594
9 0.533677219
};
\addplot+[combinedtponeh21bar,forget plot] table[x=x,y=y]{
x y
0 0.745667401
1 0.442544753
2 0.160114566
3 0.360189606
4 0.189363606
5 0.033569740
6 0
7 0.8854182758
8 0.190582786
9 0.115166958
};
\addplot+[combinedtpzeroh6bar,fill=steelblue31119180,forget plot] coordinates {(8,0.288356077783) (9,0.805774657203)};
\addplot+[combinedtpzeroh21bar,fill=steelblue31119180,forget plot] coordinates {(8,0.505048441824) (9,0.431131731383)};
\addplot+[combinedtponeh6bar,fill=turboorange,forget plot] coordinates {(8,0.643482594) (9,0.533677219)};
\addplot+[combinedtponeh21bar,fill=turboorange,forget plot] coordinates {(8,0.190582786) (9,0.115166958)};
\end{axis}
\end{tikzpicture}}%
  }
  \vspace{-0.5em}
  \caption{Comparison across traffic profiles and hours in terms of objective value, active terrestrial cells, normalized terrestrial power, and UE satisfaction. \label{fig:results_h6-h21-combined-traffic-profiles}}
  \vspace{-1.5em}
\end{figure*}

This section presents the results obtained with the proposed framework. Training and optimization were performed in one environment, while the final evaluation was conducted in a separate, unseen environment, with different random seeds used to generate the \acp{UE} and channel realizations.

For both optimization and final evaluation,
multiple \acp{UE} and channel realizations were considered,
with 10 realizations evaluated for each candidate and case.
Here we focus on two representative times of day: 06:00, corresponding to a low-load period, and 21:00, corresponding to the daily peak, for which, respectively, a total number of $103$ and $902$ \acp{UE} are considered.
Both are evaluated under low and medium traffic profiles~\cite{bernabe2026hapshypercellenablingcoverage}.
The proposed \textit{ComBO} method is compared against five baselines:
the two extreme configurations,
namely all terrestrial cells switched off (\textit{All Off}) and all cells active without  \ac{CS} (\textit{No CS}) and the
three threshold-based configurations for the Hierarchical (H) and Non-Hierarchical (NH) architectures introduced in~\cite{bernabe2026hapshypercellenablingcoverage},
and a global \ac{BO} baseline (\textit{BO}).
Results are reported in terms of objective value, number of active terrestrial cells, normalized terrestrial power consumption, as well as mean and 5th-percentile satisfaction, capturing both average \ac{QoS} and the performance of the worst-served \acp{UE}.

As shown in Figure~\ref{subfig:combined-reward},
\textit{ComBO} achieves the highest objective value in all four evaluated scenarios.
Across both traffic profiles and hours, the average objective value of \textit{ComBO} is $0.367$,
compared with $0.286$ for global  \ac{BO},
corresponding to a $28\%$ average improvement.
In general, the gains of \textit{ComBO} over global \ac{BO} range from $23\%$ to $32\%$ across all scenarios,
with the largest occurring at the daily peak under medium traffic.
These gains indicate that the proposed framework allows for identifying more effective \ac{CS} configurations that better balance consumption and \ac{QoS},
particularly under demanding traffic conditions.

Figure~\ref{subfig:combined-active-cells} shows the number of active cells per scenario. On average, \textit{ComBO} reduces the number of active terrestrial cells from $25.42$ with global \ac{BO} to $19.42$, i.e., by $24\%$.
Nevertheless, an interesting behavior can also be observed in the number and composition of active terrestrial cells.
Under low traffic,
\textit{ComBO} keeps more cells active than the most aggressive threshold-based baselines, i.e., \textit{H (0.5,0.9)} and \textit{NH (0.5,0.9)}.
The same holds against global \ac{BO} at 06:00, where \textit{ComBO} keeps $16$ cells active instead of $9$,
while, differently, at 21:00 \textit{ComBO} reduces the active cells from $22.33$ to $19$, i.e., by $15\%$.
At the same time, the optimized configurations through \textit{ComBO} tend to preserve coverage-oriented 4G cells while more selectively deactivating 5G capacity cells.
This selection suggests that \textit{ComBO} does not simply maximize \ac{CS},
but identifies a heterogeneous combination of cells that better preserves network coverage and \ac{QoS}.
In this context, the \ac{HAPS} can further support the terrestrial network by compensating for coverage gaps created by \ac{CS},
allowing capacity-oriented terrestrial cells,
which typically consume significantly more power,
to be switched off when their contribution does not justify the additional power consumption.

As a result, the advantage of \textit{ComBO} is even more pronounced in terms of terrestrial power usage, shown in Figure~\ref{subfig:combined-normalized-power}.
Averaged over the four scenarios, the normalized power $\widetilde{P}$ decreases from $0.295$ with global \ac{BO} to $0.164$ with \textit{ComBO},
corresponding to a $44\%$ reduction.
Relative to \textit{No CS}, \textit{ComBO} reduces $\widetilde{P}$ by $84\%$.

Although, in individual scenarios, some aggressive H/NH configurations achieve slightly lower $\widetilde{P}$ than \textit{ComBO},
these configurations suffer a substantial degradation in \ac{QoS}.
For example, under low traffic at 06:00,
\textit{H (0.5,0.9)} keeps only $5$ cells active, compared with $16$ for \textit{ComBO}, and achieves $\widetilde{P}$ equal to $0.10$,
while similarly, \textit{ComBO} achieves $0.11$;
however, the mean and 5th-percentile satisfaction of \textit{H (0.5,0.9)} decrease to $0.75$ and $0.15$, respectively,
while those values reach  $0.96$ and $0.81$ for \textit{ComBO}.
Hence, the additional shutdowns achieved by the greedy local policies yield only marginal power savings, but come at a significant cost in terms of \ac{UE} \ac{QoS}.


Overall, these results highlight the capability of the proposed framework to identify which cells should remain active, thereby achieving a better network operating point with a more effective trade-off between terrestrial consumption and \ac{QoS}.

Further analysis of \ac{HAPS} utilization confirms the effectiveness of the proposed optimization.
Compared with \textit{No CS},
\textit{ComBO} increases the \ac{HAPS} \ac{PRB} load from approximately $0.00$-$0.27$ to $0.18$-$0.71$,
indicating that the \ac{HAPS}-Hypercell layer effectively absorbs traffic offloaded by terrestrial \ac{CS}.
Nevertheless, the \ac{HAPS} remains underutilized,
highlighting a limitation of the current framework,
which optimizes only cell shutdown states while keeping other mechanisms fixed.
Joint optimization of association, offloading thresholds, and related mechanisms could steer more \acp{UE} toward the \ac{HAPS}-Hypercell,
thereby further exploiting its capacity and enabling additional terrestrial shutdowns.

\section{Conclusion}\label{sec:Conclusion}

In this paper, we proposed a framework for energy-efficient \ac{CS} in dense networks assisted by an \ac{HAPS}-Hypercell. By pairing terrestrial coverage and capacity cells with the Hypercell, the proposed architecture enables network-wide information aggregation, \ac{DT} modelling, and offline optimization of the \ac{CS} configuration for a target operating condition.
To address the high dimensionality of the problem, 
we proposed a scalable community-aware optimization framework that partitions the entire network into sub-groups of strongly coupled cells and optimizes them sequentially using \ac{TuRBO}. The proposed approach reduces normalized terrestrial power by $84\%$ while preserving \ac{QoS}. Compared with legacy distributed \ac{CS} baselines, it further reduces consumption by $49\%$ on average while improving satisfaction.
More broadly, the main contribution of this work lies in the proposed community-aware optimization framework. By decomposing large-scale network problems into smaller groups of strongly coupled cells, i.e., communities, the framework facilitates inclusion of additional control variables while mitigating the challenges of high-dimensional optimization. It can therefore provide a general and scalable approach to complex large network optimization problems.

\bibliographystyle{IEEEtran}
\bibliography{journalAbbreviations, bibl}


\end{document}